\documentclass[lettersize,journal]{IEEEtran}
\usepackage{amsmath,amsfonts}
\usepackage{algorithm}
\usepackage{multirow}
\usepackage{algorithmic}
\usepackage{array}
\usepackage[caption=false,font=normalsize,labelfont=sf,textfont=sf]{subfig}
\usepackage{hyperref}
\usepackage{enumitem}
\usepackage{textcomp}
\usepackage{stfloats}
\usepackage{url}
\usepackage{verbatim}
\usepackage{soul}
\usepackage{graphicx}
\usepackage{xcolor}
\usepackage{subcaption}
\counterwithin{subfigure}{figure}
\usepackage{cite}
\newcommand{\method}{\texttt{Causal-Fuzzer}}

\begin{document}

\title{
Causality-aware Safety Testing for Autonomous Driving Systems
}

\author{Wenbing Tang, 
Mingfei Cheng,
Renzhi Wang,
Yuan Zhou,~\IEEEmembership{Member,~IEEE}, 
Chengwei Liu,~\IEEEmembership{Member,~IEEE}, \\
Yang Liu,~\IEEEmembership{Senior Member,~IEEE},
and Zuohua Ding,~\IEEEmembership{Member,~IEEE}
\thanks{W. Tang, C. Liu, and Y. Liu are with the College of Computing and Data Science, Nanyang Technological University, Singapore 639798 (E-mail: 
wenbing.tang@ntu.edu.sg, chengwei.liu@ntu.edu.sg, 
yangliu@ntu.edu.sg).}
\thanks{M. Cheng is with the School of Computing and Information Systems,  Singapore Management University, Singapore 188065 (E-mail: 
mfcheng.2022@smu.edu.sg).}
\thanks{R. Wang is with the Department of Electrical and Computer Engineering,  University of Alberta, Edmonton, AB, Canada T6G 2G5 (E-mail: 
renzhi.wang@ualberta.ca).}
\thanks{Y. Zhou and Z. Ding are with the School of Computer Science and Technology, Zhejiang Sci-Tech University, Hangzhou, Zhejiang 310018, P. R. China (E-mail: yuanzhou@zstu.edu.cn, zouhuading@hotmail.com).}
}


\maketitle

\begin{abstract}
Simulation-based testing is essential for evaluating the safety of Autonomous Driving Systems (ADSs).
Comprehensive evaluation requires testing across diverse scenarios that can trigger various types of violations under different conditions. 
While existing methods typically focus on individual diversity metrics, such as input scenarios, ADS-generated motion commands, and system violations, they often fail to capture the complex interrelationships among these elements.
For instance, identical motion commands can produce different collision risks in varying scenes, and the same collision may result from different commands under different scenarios.
This oversight leads to gaps in testing coverage, potentially missing critical issues in the ADS under evaluation.
However, quantifying these interrelationships presents a significant challenge. 
In this paper, we propose a novel
causality-aware 
fuzzing technique, \method, to enable efficient and comprehensive testing of ADSs by exploring 
causally diverse 
scenarios.
The core of \method~is constructing a causal graph 
to model the interrelationships among the diversities of input scenarios, ADS motion commands, and system violations. Then the causal graph will guide the process of critical scenario generation. 
Specifically, \method~proposes (1) a causality-based feedback mechanism that quantifies the combined diversity of test scenarios by assessing whether they activate new causal relationships, and (2) a causality-driven mutation strategy that prioritizes mutations on input scenario elements with higher causal impact on ego action changes and violation occurrence, rather than treating all elements equally.
We evaluated \method~on an industry-grade ADS Apollo, with a high-fidelity simulator LGSVL.
Our empirical results demonstrate that \method~significantly outperforms existing methods in (1) identifying a greater diversity of violations (98.4 on average, compared to 42.4 for the best baseline method), (2) providing enhanced testing sufficiency with improved coverage of causal relationships (12.9 on average, compared to 6.8 for the best baseline method), and  (3) achieving greater efficiency in detecting the first critical scenarios (32.1 scenarios on average, compared to 71.5 for the best baseline method).

\end{abstract}


\section{Introduction}
\IEEEPARstart{A}{utonomous}  Driving Systems (ADSs) have made significant strides in advancing transportation intelligence and improving urban mobility. ADSs, equipped with artificial intelligence algorithms, are designed to perceive the surrounding environment through sensors such as cameras and LiDAR, and generate safe driving commands to control vehicles. Despite the rapid development of ADSs over the past few decades, the uncertainty and complexity of dynamic driving environments continue to pose safety risks~\cite{biagiola2024boundary,ji2025accelerated,chen2024misconfiguration}. Therefore, before real-world deployment, ADSs must undergo rigorous testing to identify as many potential risks as possible, ensuring their safety and reliability.

On-road testing is typically costly and impractical for covering the wide range of situations that ADSs should handle. To address this, recent studies~\cite{biagiola2024boundary,zhou2023specification,cheng2023behavexplor,huai2023doppelganger,guo2024sovar} have proposed simulation-based testing, which generates safety-critical scenarios using high-fidelity simulators, such as LGSVL~\cite{rong2020lgsvl} and CARLA~\cite{dosovitskiy2017carla}.
These methods can mainly classified into two categories: data-driven approaches~\cite{tang2023survey,zhang2023building,guo2024sovar,deng2023target} and search-based approaches~\cite{li2024viohawk,kim2022drivefuzz,cheng2023behavexplor,he2024curiosity,biagiola2024boundary,tian2022mosat}. 
Data-driven approaches aim to reconstruct safety-critical scenarios from real-world data, such as traffic accident reports~\cite{zhang2023building,guo2024sovar}, but they remain constrained by the limited availability of such data. 
In contrast, search-based approaches leverage criticality metrics (e.g., distance to collision) to guide the generation of critical scenarios, offering an effective solution for exploring safety-critical issues in ADSs. 
To minimize unnecessary exploration within the vast search space shaped by various factors (e.g., weather, traffic flow, driving behaviors), some studies~\cite{li2020av,zhou2023specification,tian2022mosat,giamattei2024causality,zhang2023accelerated} introduce advanced search algorithms, such as genetic algorithms~\cite{li2020av,zhou2023specification,tian2022mosat},
reinforcement learning~\cite{lu2022learning,haq2023many},
and surrogate models~\cite{giamattei2024causality,zhang2023accelerated}, to efficiently optimize the search process and generate safety-critical scenarios. 
Additionally, recent works~\cite{cheng2023behavexplor,huai2023sceno} have introduced extra guidance, such as the behavior diversity of the ego vehicle (i.e., the vehicle controlled by the ADS under test), to enhance the diversity of the generated scenarios during the search process.

\begin{figure*}[!t]
  \centering
\includegraphics[width=\linewidth]{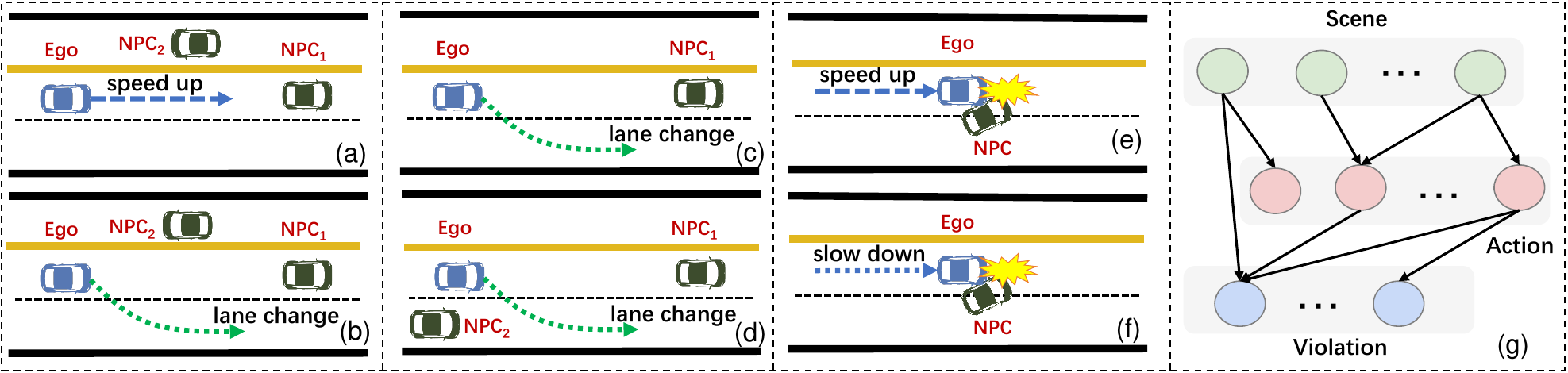}
  \caption{Diversity metrics for scenarios, actions, and violations.}
  \label{figure-motivation}
\end{figure*}

Although existing methods have studied how to generate diverse scenarios for more comprehensive testing of ADS, they consider the diversity of input scenarios~\cite{hildebrandt2023physcov,lu2022learning}, ADS actions~\cite{cheng2023behavexplor}, and violations independently.
However, they fail to explore the test case space adequately, leading to gaps in coverage that overlook diverse, critical scenarios and potential failures, ultimately resulting in insufficient testing for ADSs.
First, the diversity of input scenarios usually takes into consideration the Non-Player Character (NPC) vehicles' locations~\cite{hildebrandt2023physcov,woodlief2024s3c} and the road structures~\cite{lu2022learning}.
However, as shown in Fig.~\ref{figure-motivation}(a)-(b), even with the same road structure and NPC locations, different behaviors taken by the ADS can lead to varied results, where accelerating may result in a collision while lane-changing could allow for safe passage.
\emph{Therefore, it is insufficient to consider only the diversity of the input space.}
Second, 
existing methods typically assess the behavioral diversity of the ego vehicle using action-related elements such as its position, velocity, and acceleration.
For example, according to existing action diversity~\cite{cheng2023behavexplor}, the two scenarios in Fig.~\ref{figure-motivation}(c)-(d) will be classified as one category, as they both involve the same action of lane-changing.
However, Fig.~\ref{figure-motivation}(d) considers target lane occupancy, where the ego vehicle is at risk of colliding with NPC$_2$, whereas the ego vehicle remains safe in Fig.~\ref{figure-motivation}(c).
\emph{Therefore, relying solely on action diversity can overlook critical distinctions of scenarios, resulting in missing certain critical scenarios.}
Third, other works consider the diversity of violations, such as distinguishing between vehicle-vehicle and vehicle-pedestrian collisions~\cite{huai2023doppelganger,guo2024sovar}.
\emph{However, a violation can result from various driving conditions and actions. Failing to analyze the conditions under which the violation occurs may overlook distinct critical scenarios. 
}
For instance, before the collision in Fig.\ref{figure-motivation}(e), the ego vehicle was accelerating, which is clearly erroneous, while in Fig.\ref{figure-motivation}(f), the ego vehicle is attempting to slow down.
Therefore, it is crucial to develop a novel testing approach that considers the diversity of input scenarios, actions, and violations, and their interrelationships to achieve comprehensive ADS testing.

However, characterizing the interrelationships of different diversity metrics poses several challenges.
First, given the complex traffic flow with temporal motion and spatial distribution of vehicles, it is challenging to represent scenarios, actions, and violations in a unified format.
Second, in open and unpredictable environments, the number of possible scenarios is vast, potentially infinite. Additionally, the range of actions the ego vehicle can take is both complex and limitless, and so are potential system violations. Consequently, the interrelationships among scenarios, actions, and violations lead to a state explosion problem, making it challenging to analyze the interrelationships.
Third, the interrelationships among scenarios, actions, and violations are often non-linear. Capturing such non-linear interrelationships is difficult and requires advanced methods.
Finally, irrelevant data is often present in a scenario. 
For example,  as shown in Fig.~\ref{figure-motivation}(a)-(b), NPC$_2$, driving in the opposite lane, does not influence the ego vehicle's actions.
Consequently, when calculating the interrelationships, it is crucial to filter out such irrelevant information while retaining the most essential semantic information.

To tackle these challenges, we propose \method, a causality-aware testing framework that efficiently generates diverse critical scenarios. Our approach measures diversity through the lens of causality, enabling the characterization of the interrelationships between different diversity metrics.
Specifically, \method~designs a causality-based feedback that extracts a \textit{scene-action-violation} causal graph from data collected at individual time snapshots (i.e., scenes) during scenario execution, as shown in Fig.~\ref{figure-motivation}(g).
Based on the causal graph, we define a violation diversity metric that calculates the distance between the edge combinations of $(scene, action) \rightarrow violation$ in a new test case and those in existing test cases.
Additionally, we define a testing sufficiency metric that measures the distance over edge combinations of $scene \rightarrow action$.
An effective testing process should aim to maximize violation diversity while ensuring high testing sufficiency.
Furthermore, based on the extracted causal graph, we propose a causality-driven adaptive mutation strategy that estimates each NPC's causal effects on changes in ego actions and the occurrence of violations.
The strategy then prioritizes mutations for NPCs with  stronger causal effects, while NPCs with weaker causal effects, such as  NPC$_2$ in Fig.~\ref{figure-motivation}(a), 
are mutated with a lower probability.
By leveraging an understanding of causal relationships among different objects, this mutation strategy improves efficiency while maintaining interpretability.

We evaluate the effectiveness and efficiency of \method~on the LGSVL simulator and Baidu Apollo~\cite{apollo}.
Specifically, we first conduct experiments to verify whether the causal graph can precisely represent a scenario.
The results show that the extracted causal graph correctly captures the semantic information within the scenario.
Then, we compare \method~with three baselines, i.e., random testing, AV-Fuzzer~\cite{li2020av}, and DoppelTest~\cite{huai2023doppelganger}.
The evaluation results demonstrate that \method~can discover more safety violations, achieving both high violation diversity and strong testing sufficiency.
Furthermore, we analyze the efficiency of \method, and the results indicate that our method is more efficient and requires less exploration to identify the first critical scenario.
Finally, we conduct ablation studies to justify the usefulness of 
causality-based feedback and causality-driven mutation.

The main contributions of this paper are as follows:
\begin{itemize}[leftmargin=*]
\item We propose a novel causality-aware fuzzing method, \method, to discover critical scenarios that achieve both high violation diversity and strong testing sufficiency.
\item We propose a causality-based feedback mechanism to effectively and efficiently guide the search for diverse violations.
\item We propose a causality-driven mutation strategy that incorporates causal information to enhance mutation efficiency in an interpretable manner.
\item We evaluate \method~on Baidu Apollo, an industry-grade ADS, and the results show that \method~improve testing sufficiency by 22.06\% and increases violation diversity by 134.09\%.
\end{itemize}


The rest of this paper is organized as follows.
Section~\ref{preliminaries-background} presents the background and preliminaries.
The detailed procedures of \method~are provided in 
Section~\ref{approach-details}.
Experiments are conducted in Section~\ref{experiment-sec} to demonstrate the  effectiveness and efficiency of our approach.
Section~\ref{related-work-sec} summarizes the related work.
Finally, conclusion and future work are discussed in Section~\ref{conclusion-sec}.

\section{Background}
\label{preliminaries-background}
\subsection{Preliminaries on ADS Testing}
\label{preliminaries-ADS}
\textbf{Autonomous Driving Systems (ADSs):} An ADS is designed to achieve high automation levels for vehicles
to automatically drive on the roads.
Existing ADSs can be divided into two categories: (1) end-to-end driving models, and (2) multi-module driving models.
End-to-end models consider the entire driving system as a neural network model from receiving sensing data to generating control commands.
Although end-to-end models have made significant advancements in recent years~\cite{chen2024end,shao2024lmdrive,doll2024dualad,hu2023planning}, they still encounter challenges such as heavy reliance on data quality, limited interpretability, and poor generalizability. As a result, end-to-end models have yet to achieve widespread adoption in industrial fields.
In contrast, multi-module ADSs (e.g., Baidu Apollo~\cite{Apollo2019} and Autoware~\cite{Autoware2022})
are composed of various modules~\cite{zhou2023specification,feng2024rocas}, including localization, perception, prediction, planning, and control.
The localization module determines the vehicle’s state information.
The perception module interprets sensor data to understand the surrounding environment, such as obstacle object recognition and traffic signal identification.
The prediction module anticipates the future trajectories of the surrounding objects.
Given the results of the perception and prediction modules, planning
is responsible for generating a collision-free trajectory for the vehicle.
Finally, the control module generates proper control commands (i.e., steering, throttle, brake) for the vehicle chassis.
Due to their reliable performance and good interpretability, multi-module ADSs have gained widespread adoption in the industry.
Therefore, we focus on multi-module ADS testing in this work.

\textbf{Scenario and Scene:} 
ADS testing aims to identify critical scenarios that expose safety violations in the ADS under test. A scenario describes the relevant characteristics, activities, and/or
goals of the ego vehicle and other objects (e.g., pedestrians and NPCs).
A scenario can be described as a sequence of scenes, where a scene can be regarded as a snapshot of the world.
According to \cite{menzel2018scenarios}, scenarios can be classified into three types: functional scenarios, which describe roads and objects using linguistic notation; logical scenarios, which define the parameters and their ranges within a functional scenario; and concrete scenarios, which assign specific values to each of the parameters outlined in a logical scenario.
A concrete scenario can be executed in a simulator.
Hence, we need to search for critical concrete scenarios from the huge scenario space.

\textbf{Scenario Observation:} 
In ADS testing, a complex concrete scenario can consist of hundreds or even thousands of attributes extracted from the Operational Design Domains (ODDs)~\cite{thorn2018framework}.
Given the safety
violations, we focus on the configurable attributes of each NPC vehicle to search for critical scenarios.
Hence, we mutate the NPC's route and its corresponding waypoints (i.e., position and velocity) in each scenario.
After executing a scenario $s$ with a running duration of $t^s$, we can obtain the corresponding observation $\boldsymbol{O}^s$.
Since $t^s$ can be divided into a set of scenes (i.e., discrete-time instants)~\cite{ulbrich2015defining}, the scenario observation can be denoted as $\boldsymbol{O}^s = \{\langle \boldsymbol{oe}_t^s, \boldsymbol{on}_t^s \rangle|t = 1, 2, \cdots, t^s\}$.
At any time instant $t$, the ego vehicle's state can be represented as $\boldsymbol{oe}_t^s = \langle t, (p_t^0, h_t^0, v_t^0, a_t^0) \rangle$, where $p_t^0$, $h_t^0$, $v_t^0$, and $a_t^0$ denote the position, heading, velocity, and acceleration of the ego vehicle, respectively. Additionally, the states of the NPC vehicles are denoted as  $\boldsymbol{on}_t^s = \{\langle t, (p_t^k, h_t^k, v_t^k, a_t^k) \rangle | k = 1, 2, \cdots, n^{NPC}\}$, where $p_t^k$, $h_t^k$, $v_t^k$, and $a_t^k$ represent the position, heading, velocity, and acceleration of NPC vehicle $k$, respectively; $n^{NPC}$ denotes the number of NPC vehicles.
The states of the ego vehicle and NPC vehicles at a time instant $t$ form the scene of the scenario at $t$.

\textbf{Safety Specifications:} ADSs should satisfy various specifications. In this paper, we choose the essential safety and task achievement specifications.
We check whether the ego vehicle successfully
reaches its destination without encountering any collisions with NPC vehicles.
Given a scenario $s$ and its execution observation $\textbf{O}^s$, the test oracle’s passing condition (i.e., not triggering a safety violation) can be defined as follows:
\begin{equation}
D_{p2p}(p_{t^s}^0, p_{dest}) \le \theta_p
\bigwedge D_b(p^0, s) > 0.
\label{eq-safety-violation-sepecification}
\end{equation}
where $D_{p2p}(p_{t^s}^0, p_{dest})$ calculates the distance between the last position of the ego vehicle, i.e., $p_{t^s}^0$, and the destination $p_{dest}$. Note that there will be some estimation errors in the simulator (e.g. length of vehicles), we consider the ego vehicle to have successfully reached its destination if the distance falls within a specified threshold $\theta_p$.
$D_b(p^0, s)$ denotes the minimum distance between the ego vehicle and the NPC vehicles in scenario $s$ during the execution, i.e., 
\begin{equation}
D_b(p^0, s) = \min \left(
\{D_{b2b}(p_t^0, p_t^k)|1 \le t \le t^s, 1 \le k \le n^{NPC}\}
\right).
\label{minimum-distance-calculation}
\end{equation}
where $D_{b2b}(p_t^0, p_t^k)$ computes the shortest distance between the
bounding boxes of the ego vehicle and NPC vehicle $k$ at the time instant $t$, whose positions are $p_t^0$
 and $p_t^k$, respectively.

\subsection{Preliminaries on Causality}

\textbf{Causal Graph:} This paper measures the diversity through the lens of causality. 
The causality among a set of variables within a system can be represented by a causal graph $\mathcal{G} = (\mathcal{V}, \mathcal{E})$, where $\mathcal{V}$ stands for a set of variables (nodes) in the graph, and $\mathcal{E} \subseteq \mathcal{V} \times \mathcal{V}$ denotes the causal relationships (directed edges) among those variables~\cite{pearl2009causality}.
Actually, each causal graph can be represented graphically as a Directed Acyclic Graph (DAG).
In particular, causal graphs allow us to distinguish cause and effect
based on whether a node is an ancestor or descendant of another node.
For any two nodes $V_i, V_j \in \mathcal{V}$,
the presence/absence of an edge $V_i \rightarrow V_j$ represents the presence/absence of a direct causal effect of $V_i$ on $V_j$.
Hence, the edges in a causal graph can help us understand the data-generating process and functional mechanism of a system~\cite{xia2024deciphering,sun2024neural}.
However, in real-world applications, the causal graph of the underlying system is often unknown. 
Instead, we have to collect a set of observational data from the target system to discover the causal graph 
$\mathcal{G}$.

\textbf{Causal Effect:} 
Besides identifying the causal relationships among variables, we can further quantify the strengths (i.e., causal effects) of these causal relationships.
Given two nodes $V_i$ and $V_j$ in a causal graph $\mathcal{G}$,
the causal effect of $V_i$  on $V_j$ quantifies the change in the outcome $V_j$ brought from the change in a cause $V_i$.
For example, let $V_i$ represent ``\textit{whether there is an NPC on the left side of the ego vehicle}'', and $V_j$ denote ``\textit{whether a collision has been detected}''. In this case, the causal effect of $V_i$ on $V_j$ indicates ``\textit{the causal effect of the presence of a vehicle on the left side on the occurrence of a collision}''.
 In such case, $V_i$ is a binary variable and the interested Average Causal Effect (ACE) can be described as:
 \begin{equation}
ACE(V_i \rightarrow V_j) = \mathbb{E}[V_j|do(V_i = 1)] -  \mathbb{E}[V_j|do(V_i = 0)].
\end{equation}
where the \textit{do-calculus} is an intervention operation~\cite{pearl2009causality}, setting a variable to a constant value. ACE refers to the expected change in the average of the outcomes.
Hence, variable $V_i$ has a causal effect on variable $V_j$ if and only if $\mathbb{E}[V_j|do(V_i = 1)]  \ne \mathbb{E}[V_j|do(V_i = 0)]$.
In general, if variables $V_i$ and $V_j$ can each take on more than one value, we need to estimate the
effect $\mathbb{E}[V_j=v_j|do(V_i = v_i)]$, where $v_i$ and $v_j$ are any two values that $V_i$ and $V_j$ can take on.

\section{Approach}
\label{approach-details}
\subsection{Overview}
\begin{figure*}[!t]
  \centering
\includegraphics[width=\linewidth]{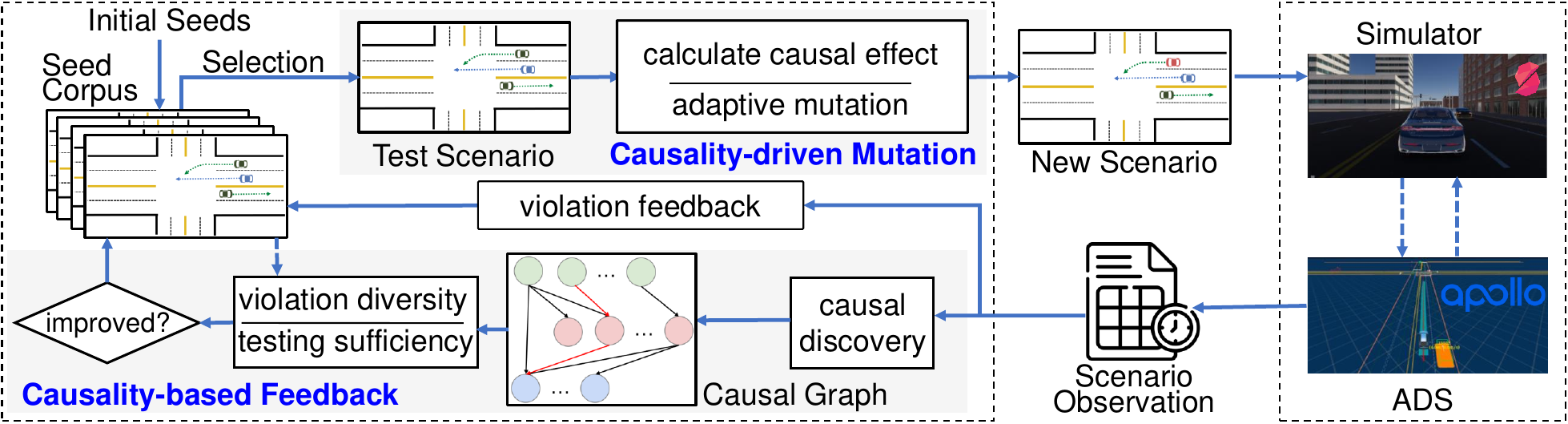}
  \caption{The framework of \method. It contains a causal-driven mutation strategy and a causality-based feedback method.}
  \label{figure-framework}
\end{figure*}

In this paper, we propose  \method~to generate diverse critical scenarios for ADS, which aims at not only criticality but also diversity and efficiency.
Figure~\ref{figure-framework} illustrates the framework of \method.
\method~maintains a corpus of seeds, each representing a scenario. 
At each iteration, it first selects a seed from the corpus, e.g., the seed with the best fitness.
Second, a \textit{causality-driven mutation} strategy is proposed to generate new test cases by mutating the selected seed.
The generated test case is then executed on the simulation platform. 
Finally, a \textit{causality-based feedback} is 
employed to assess whether safety violations occurred and whether the violation diversity and testing sufficiency metrics are improved.

In the feedback phase, \method~leverages scenario criticality, violation diversity, and testing sufficiency as the fuzzing feedback to guide the search for ``interesting" test cases.
Specifically, safety violation functions are first defined to evaluate the criticality of the mutant, i.e., the violation degree of the safety specifications in a scenario.
In addition, upon receiving a scenario observation, \method~conducts a causal discovery process to extract causal graphs that represent the interrelationships among scenes, actions, and violations within the scenario.
A diversity evaluation process then quantifies the differences between the extracted causal graphs and those of existing seeds by computing the distances between their adjacency matrices.
Hence, for each seed in the corpus, we maintain both the seed and its corresponding causal graph. 
By incorporating the causal graph, we capture the essential interrelationships within a scenario, providing a way to quantitatively measure diversity.
Mutants with higher violation degrees, greater diversity, or improved testing sufficiency are added to the seed corpus.

In the mutation phase, a seed is first selected from the input corpus based on its fitness.
Then, \method~calculates the causal effect of each NPC vehicle on the changes of ego actions and the occurrence of violation.
The causal effect quantifies the overall contributions of each NPC vehicle to violation diversity and testing sufficiency. 
This measure helps in identifying which NPC vehicles are more likely to influence the outcomes of a scenario.
Hence, \method~adopts an adaptive mutation strategy, prioritizing the mutation for these influential NPC vehicles with higher probabilities. 

The algorithm of \method~is given in Algorithm~\ref{alg1}.
It takes as input a set of initial seeds $\Delta$ and the ADS under test, along with configurable parameters such as the testing sufficiency threshold and violation diversity threshold. 
The algorithm's outputs include a set of useful scenarios $\Delta$, the failed tests $\boldsymbol{FT}$, unique combinations of \textit{scene-action} causal edges $\boldsymbol{SAC}$ that indicate testing sufficiency, and distinct combinations of \textit{scene-action-violation}  causal edges $\boldsymbol{SAVC}$ that represent causally different test scenarios.
Lines~\ref{alg-init-start}--\ref{alg-init-end} describe the initialization step.
\method~begins with initializing three empty sets  $\boldsymbol{FT}$, $\boldsymbol{SAC}$, $\boldsymbol{SAVC}$ (Line~\ref{three-sets-initialization}).
For each seed in the initial corpus, it is executed in the simulator with the ADS.
After each execution, the scenario observation $\boldsymbol{O}^s$, the execution result $R^s$, and the fitness value $Fit^s$ are collected (Line~\ref{init-seed-execution}).
Then, a \textit{scene-action-violation} causal graph $\mathcal{G}^{s}_{sav}$,
is discovery based on  the scenario observation $\boldsymbol{O}^s$ and the 
execution result
$R^s$
(Line~\ref{init-causal-discovery}).

\begin{algorithm}[h]
\caption{\method~Testing Algorithm.}
\label{alg1}
\textbf{Input}: Initial seed corpus $\Delta$, target ADS system, testing sufficiency threshold $\theta_{ts}$, violation diversity  threshold $\theta_{vd}$.\\
\textbf{Output}: Useful seeds $\Delta$, failed test set $\boldsymbol{FT}$, \textit{scene-action} combinations $\boldsymbol{SAC}$, \textit{scene-action-violation} combinations $\boldsymbol{SAVC}$.
\begin{algorithmic}[1]
    \STATE $\boldsymbol{FT} \leftarrow \emptyset$, $\boldsymbol{SVC} \leftarrow \emptyset$, $\boldsymbol{SAVC} \leftarrow \emptyset; $\label{alg-init-start} \label{three-sets-initialization}
    \FOR{$s \in \Delta$} 
    \STATE $\boldsymbol{O}^s, R^s, Fit^s \leftarrow \mathtt{Simulator}(s;  ADS)$; \label{init-seed-execution}
    \STATE $\mathcal{G}^s_{sav}  \leftarrow \mathtt{CausalDiscovery}(\boldsymbol{O}^s, R^s)$; \label{init-causal-discovery}
    \ENDFOR \label{alg-init-end}
    \WHILE{$runnign~time \leq budget$} \label{fuzzing-loop-start}
    \STATE $s \leftarrow \mathtt{SeedSelection}(\Delta)$; \label{alg-seed-selection}
    \STATE $s^\prime \leftarrow \mathtt{CausalAdaptiveMutation}(s; \mathcal{G}^s_{sav})$;\label{alg-seed-mutation}
    \STATE  $\boldsymbol{O}^{s^\prime}, R^{s^\prime}, Fit^{s^\prime} \leftarrow \mathtt{Simulator}(s^\prime; ADS)$; \label{alg-seed-execution}
    \STATE $\mathcal{G}^{s^\prime}_{sav}  \leftarrow \mathtt{CausalDiscovery}(\boldsymbol{O}^{s^\prime}, R^{s^\prime})$; \label{alg-fuzzing-causal-doscovery}
    \IF{$R^{s^\prime} = violated$} \label{alg-fuzzing-violated-start}
    \STATE $\boldsymbol{FT} \leftarrow \boldsymbol{FT} \cup \{s^\prime\}, \Delta \leftarrow \Delta \cup \{s^\prime\}$;
     \label{alg-fuzzing-violated-end}
     \ELSE
     \IF{$\mathtt{Distance}(\mathcal{G}^{s^\prime}_{sav}, \boldsymbol{SAC}) \geq \theta_{ts}$ and $Fit^{s^\prime} < Fit^s$} 
     \label{alg-behavior-set-start}
     \STATE $\boldsymbol{SAC} \leftarrow \boldsymbol{SAC} \cup \{s^\prime\}, \Delta \leftarrow \Delta \cup \{s^\prime\}$;
     \label{alg-behavior-set-end}
     \ELSE
     \IF{$\mathtt{Distance}(\mathcal{G}^{s^\prime}_{saf}, \boldsymbol{FP}) \geq \theta_{vd}$} \label{alg-pattern-set-start}
     \STATE $\boldsymbol{SAVC} \leftarrow \boldsymbol{FP} \cup \{s^\prime\}, \Delta \leftarrow \Delta \cup \{s^\prime\}$;
     \ENDIF
\ENDIF
\ENDIF
     \label{alg-pattern-set-end}
    \ENDWHILE \label{fuzzing-loop-end}
    \RETURN $\Delta$, $\boldsymbol{FT}$, $\boldsymbol{SAC}$, and $\boldsymbol{SAVC}$.\label{alg-return-line}
\end{algorithmic}
\end{algorithm}

In the algorithm, during the fuzzing procedure~(Lines~\ref{fuzzing-loop-start}--\ref{fuzzing-loop-end}), 
\method~repeats the mutation, execution, and feedback processes iteratively until the stopping criterion is satisfied, e.g., when the given time budget expires.
In each iteration, a seed 
$s$ is first selected from the corpus (Line~\ref{alg-seed-selection}), and it is mutated to generate a new seed $s^\prime$ through the causality-driven mutation strategy (Line~\ref{alg-seed-mutation}).
Then, the new mutant $s^\prime$ is executed in the simulator, 
and its execution results, including
$\boldsymbol{O}^{s^\prime}, R^{s^\prime}$, and $Fit^{s^\prime}$, are collected (Line~\ref{alg-seed-execution}).
In Line~\ref{alg-fuzzing-causal-doscovery}, based on the current received observation $\boldsymbol{O}^{s^\prime}$ and its execution result $R^{s^\prime}$, \method~discovers a
new causal graph $\mathcal{G}^{s^\prime}_{sav}$, encoding the causal relationships activated during the execution of mutant $s^\prime$.
If the mutant leads to a violation, it is added to the failed test set 
$\boldsymbol{FT}$ and the seed corpus $\Delta$ (Lines~\ref{alg-fuzzing-violated-start}--\ref{alg-fuzzing-violated-end}).
A mutant is added to the set 
$\boldsymbol{SAC}$
and the seed corpus $\Delta$ if it simultaneously satisfies two conditions (Lines~\ref{alg-behavior-set-start}--\ref{alg-behavior-set-end}): (1) 
testing sufficiency (i.e., the distance between the current \textit{scene-action} causal graph
$\mathcal{G}^{s^\prime}_{sav}$ and the existing graphs in  $\boldsymbol{SAC}$)
exceeds a predefined threshold $\theta_{ts}$, and (2) the violation degree of $s^\prime$ is less than that of its parent.
Note that these two conditions ensure that our method enhances both testing sufficiency and violation degree, rather than merely increasing sufficiency.
Moreover,
if the violation diversity improves, meaning that the distance between the current \textit{scene-action-violation} causal graph $\mathcal{G}^{s^\prime}_{sav}$
  and the existing graphs in 
$\boldsymbol{SAVC}$ exceeds a given threshold 
$\theta_{vd}$, 
the mutant $s^\prime$ will be added to both the set $\boldsymbol{SAVC}$ and the seed corpus $\Delta$ (Lines~\ref{alg-pattern-set-start}--\ref{alg-pattern-set-end}).
The three types of feedback work together, enabling \method~to discover a greater diversity of violations while simultaneously facilitating more comprehensive testing.
Finally, the algorithm returns $\Delta$, $\boldsymbol{FT}$, $\boldsymbol{SAC}$, and $\boldsymbol{SAVC}$ (Line~\ref{alg-return-line}).

\subsection{Causality-based Feedback}
\label{causal-feedback}

\method~aims to discover diverse violations while ensuring sufficient testing.
It incorporates feedback from three metrics: \textit{violation diversity}, \textit{testing sufficiency}, and \textit{violation degree}.
Thus, a key challenge lies in calculating these metrics.
To this end, we first abstract the observational data of a scenario into a vectorized representation (Section~\ref{scenario-abstraction}).
This representation is then used to construct the causal graph (Section~\ref{causal-graph-discovery}).
Finally, feedback is calculated based on the causal graph (Section~\ref{feedback-computation}).

\subsubsection{Scenario Observation Vectorization}
\label{scenario-abstraction}
Given a scenario $s$ that is fed into the target ADS, we can obtain an observation trace $T^s = \langle\textbf{O}^s, \Phi^s \rangle = \{\langle \textbf{oe}_t^s, \textbf{on}_t^s, \phi^s_t \rangle|t = 1, 2, \cdots, t^s\}$, where $\textbf{oe}_t^s$ and  $\textbf{on}_t^s$ represents
the states of the ego vehicle and the NPC vehicles at $t$, respectively, and $\phi^s_t$ indicates whether a violation has been detected at scene $t$.
Given the high-dimensional and variable nature of scenes, we first abstract the observation trace $T^s$ into a vectorized scenario representation $\textbf{X}^s$ before constructing causal graphs.
However, there are two significant challenges in the vectorization process: (1) 
due to the varying numbers of NPC vehicles in different scenarios, the length of $\textbf{on}_t^s$ will differ, which poses a challenge for unifying the length of the vector representation $\textbf{X}^s$; and (2) 
since the observation trace $T^s$ contains both discrete variables (e.g., the failure indicator $\phi^s_t$) and continuous variables (e.g., the position and velocity in $\textbf{oe}_t^s$ and $\textbf{on}_t^s$), the vectorization process becomes more complex, as the heterogeneous data types present challenges for constructing causal graphs.
In the following, we provide the details of the abstraction process, which consists of four components: scene abstraction, action abstraction, violation abstraction, and scenario abstraction.

\begin{figure*}[!t]
  \centering
\includegraphics[width=\linewidth]{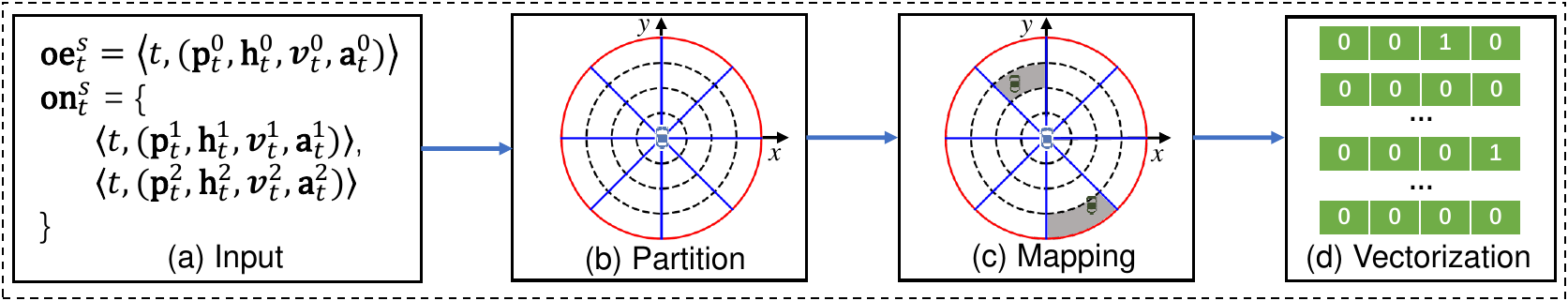}
  \caption{The vectorization process of a scene within a test scenario.}
  \label{figure-scene-abstraction}
\end{figure*}

\textbf{Scene Abstraction:} 
For each time instant, we first generate the scene abstraction, which is the abstraction of the ADS's inputs.
At scene $t$, suppose the state of the ego vehicle is $\textbf{oe}_t^s = \langle t, (p_t^0, h_t^0, v_t^0, a_t^0) \rangle$, which records ADS position $p_t^0$, heading $h_t^0$, velocity $v_t^0$, and acceleration $a_t^0$. Additionally, the state of all NPC vehicles is given by $\textbf{on}_t^s = \{\langle t, (p_t^k, h_t^k, v_t^k, a_t^k) \rangle | k = 1, 2, \cdots, n^{NPC}\}$, where $p_t^k$, $h_t^k$, $v_t^k$, and $a_t^k$ represent the position, heading, velocity, and acceleration of the NPC vehicle $k$.
Hence, we propose an abstraction approach to transform the observation into a vectorized 
 scene representation $\textbf{sr}_t^s$.
The detailed vectorization process is shown in Fig.~\ref{figure-scene-abstraction}.

Given that modern autonomous vehicles typically utilize LiDAR with a 360-degree field of view to perceive their surrounding environments~\cite{hildebrandt2023physcov,cao2023you}, we first define a perception region $PR$, represented by the red circle in Fig.~\ref{figure-scene-abstraction}(b).
In that case, NPC vehicles within the region are more likely to affect the ego vehicle's behavior, while those located outside the region are considered less relevant.
Hence, for time instant $t$ in scenario $s$, 
we define a local Cartesian coordinate system
whose origin is the center of $p_t^0$, the y-axis is the same as the ego heading $h_t^0$, and the \textit{x}-axis is perpendicular to the
\textit{y}-axis.
Therefore, the perception region can be denoted as 
$PR_t^s = \{p \in \mathbb{R}^2|D_{p2p}(p, p_t^0) \leq L\}$, where $L$ is the perception range.
Furthermore, we partition region $PR_t^s$ into  $m$ equal
sectors, marked with blue lines in Fig.~\ref{figure-scene-abstraction}(b),
i.e., $PR_t^s = \bigcup_{i=1}^{i=m} pr_i$, where $pr_i = \{(x, y)|0 \leq x  \leq L \cos(\theta_i), 0 \leq y \leq L \sin(\theta_i), 2(i-1)\pi/m \leq \theta_i \leq 2i\pi/m\}$.
For each sector, it is further  partitioned into $n$ annular sectors, represented with dashed lines in Fig.~\ref{figure-scene-abstraction}(b),
i.e., $pr_i = \bigcup_{j=1}^{j=n} as_j$, where $as_j =
\{(x, y)|(j-1)L/n \cos(\theta_i) \leq x  \leq jL/n \cos(\theta_i), (j-1)L/n\sin(\theta_i) \leq y \leq jL/n \sin(\theta_i)\}$.
Consequently, the perception region is partitioned into $mn$ disjoint areas.

In order to generate a uniform-length scene representation, the next step is to map the NPC vehicles into the partition, ensuring that the varying numbers of NPC vehicles across different scenarios can be effectively encoded into a consistent-length vector structure.
NPC vehicles outside the perception region will not be perceived by the ego vehicle due to their distance, and therefore will not be represented at the current time.
However, they will be characterized once they enter the ego vehicle's perception region.
For NPC vehicles located within  $PR$, we transform their positions into the local Cartesian coordinate system, as shown in Fig.~\ref{figure-scene-abstraction}(c).
For each area in $PR$, we use a binary variable
to indicate whether there are any NPC vehicles present within the area.
Specifically, the indicator variable for a given area will be assigned a value of $1$ if one or more NPC vehicles are present within that area; otherwise, it will be assigned a value of $0$.
For example, as shown in Fig.~\ref{figure-scene-abstraction}(c), suppose $m = 8$ and $n = 4$. In this case, there is an NPC vehicle positioned in the third annular sector to the left front of the ego vehicle; therefore, the representation vector for this area is $[0,0,1,0]$.
Similarly, in the right rear area, an NPC vehicle is located in the fourth annular sector, resulting in a vector representation of $[0,0,0,1]$.
Note that an NPC vehicle may simultaneously occupy multiple areas. In this way, the region assignment is determined based on the geometric center of the vehicle, ensuring a consistent and unambiguous representation in the vectorization process.

Finally, by concatenating the representation vectors of each area, we obtain a vectorized 
scene representation $\boldsymbol{sr}_t^s$ of length $mn$, where each element at a given position indicates the occupancy status of the corresponding area by NPC vehicles.
This representation effectively captures the spatial distribution of NPC vehicles while mitigating the effects of varying numbers of NPCs on the length of the representation vector, thus ensuring a uniform-length representation. 
Consequently, this abstraction approach is suitable for scenes involving any number of NPC vehicles.

\textbf{Action Abstraction:} 
In order to obtain vectorized 
 action representation $\textbf{ar}_t^s$, we define five indicator variables as: $\textbf{ar}_t^s = [aiv_{+}, aiv_{-}, aiv_{l},
 aiv_{r}, aiv_{m}]$, 
where variables $aiv_{+}$, $aiv_{-}$, $aiv_{l}$, $aiv_{r}$, and $aiv_{m}$ are binary ($0$ or $1$),
indicating the absence or presence of the corresponding  acceleration, deceleration, left turn, right turn, and maintaining behaviors of the ego vehicle.
Specifically, variables $aiv_{+}$ and $aiv_{-}$
are determined by the current acceleration $a_t^0$, as follows:
\[
\begin{cases}
    aiv_{+} = 1~\text{if } a_t^0 \geq \theta_a^{+},
    & aiv_{+} = 0~  \text{otherwise}; \\
    aiv_{-} = 1~\text{if } a_t^0 \leq \theta_a^{-}, 
    & aiv_{-} = 0~ \text{otherwise}.
\end{cases}
\]
where $\theta_a^{+}$ and $\theta_a^{-}$ are two thresholds used to determine whether acceleration and deceleration events have occurred.
The variables $aiv_{l}$ and $aiv_{r}$ are defined based on the difference between the current heading angle $h_t^0$ and the previous heading angle $h_{t-1}^0$:

\[
\begin{cases}
    aiv_{l} = 1 \text{ if } h_{t-1}^0 - h_t^0 \geq \theta_h,
     &  aiv_{l} = 0 \text{ otherwise};\\
    aiv_{r} = 1  \text{ if } h_t^0 - h_{t-1}^0 \geq \theta_h,  & 
    aiv_{r} = 1 \text{ otherwise}.
\end{cases}
\]
where $\theta_h$ is a threshold to identify significant turning events while disregarding minor directional changes.
The variable $aiv_{m}$ is assigned a value of 1 only when the ego vehicle is exhibiting uniform linear motion, which occurs when both its acceleration and changes in heading are below their respective defined thresholds.
Note that for $t=0$,  we initialize the vector as $\textbf{ar}_t^s = [0, 0, 0, 0, 0]$.
For $t \geq 1$, we calculate 
$\textbf{ar}_t^s$
based on the current and previous states of the ego vehicle, i.e.,  $\textbf{oe}_t^s$ and $\textbf{oe}_{t-1}^s$.
For example, if we obtain an action representation 
 $\textbf{ar}_t^s = [1, 0, 0, 1, 0]$, it indicates that the ego vehicle is currently accelerating while making a right turn, which could occur during a lane change for overtaking.
By the above action abstraction, we can extract the semantic information from the ego's actions and encode it.

\textbf{Violation Abstraction:} The next abstraction step is dedicated to generating the vectorized violation representation $\textbf{vr}_t^s$ based on the violation indicator $\phi^s_t$.
We introduce a variable $f_e$  to indicate whether the ego vehicle has caused a collision.
Additionally, we define a variable $f_n$ to signify whether a collision has been caused by an NPC vehicle.
Specifically, the collision is monitored by checking the minimum distance between the ego vehicle and the NPC vehicles, as defined in~\eqref{eq-safety-violation-sepecification}.
Following existing works~\cite{li2020av,huai2023doppelganger}, whether the collision is caused by the ego vehicle or by the NPC vehicles is determined based on their current speeds and whether any lane-crossing behavior has occurred.
For example, when $\textbf{vr}_t^s = [1,0]$, it indicates that a violation caused by the ego vehicle occurred at the current time instant.

\textbf{Scenario Abstraction:} 
Consequently, by integrating the scene representation $\textbf{sr}_t^s$, the action representation $\textbf{ar}_t^s$, and the violation representation  $\textbf{vr}_t^s$, we can generate a comprehensive representation vector, i.e.,  $[\textbf{sr}_t^s, \textbf{ar}_t^s, \textbf{vr}_t^s]$, for the observation at time instant $t$.
However, the aforementioned approach only describes the abstraction process at each time instant $t$.
For a given scenario $s$, we need to repeat the abstraction process $t^s$ times to produce a scenario-level representation $\textbf{X}^s$. 



\subsubsection{Causal Graph Discovery}
\label{causal-graph-discovery}

For each scenario, we can collect its vectorized representation  
$\boldsymbol{X} = [x_1, x_2, \cdots, x_u] \in \mathbb{R}^{u \times q}$, where $u$ represents the dimension of the representation vectors, and $q$ is the length of the scenario observation.
The goal of causal discovery is to infer causal relationships among variables from the observational data in a data-driven manner, based on the premise that causality can be identified through statistical dependencies~\cite{giamattei2024causality,zanga2022survey}.
The task can be described as:
 \begin{equation}
\min_{\boldsymbol{W} \in \mathbb{R}^{u \times u}} \mathcal{S}(\boldsymbol{W}, \textbf{X})~s.t.~\mathcal{G}(\boldsymbol{W})~ being~a~DAG.
\label{causal-discovery-problem}
\end{equation}
where $\mathcal{S}(\cdot)$ is a consistency score that measures the consistency between the probability distribution associated with the causal graph and the observational data.
The matrix $\boldsymbol{W} = {w_{ij}} \in \mathbb{R}^{u \times u}$ is a weighted adjacency matrix, where $w_{ij}$ represents the causal strength from variable $x_j$ to variable $x_i$ in the DAG.
Therefore, our goal is to determine the adjacency matrix $\boldsymbol{W}$ based on the observational data $\boldsymbol{X}$.

In this paper, to solve the problem defined in~\eqref{causal-discovery-problem}, we use the Linear Non-Gaussian Acyclic Model (LiNGAM)~\cite{shimizu2006linear}, which has been proven can efficiently identify a much broader range of causal relationships than classical causal discovery methods~\cite{shimizu2022statistical,ikeuchi2023python}.
In LiNGAM, for a dataset with $u$ observed variables, its data generation process can be formalized
 as:
\begin{equation}
x_i = \sum_{x_j \in pa(x_i)} w_{ij} x_j + e_i~ (i=1,\cdots,u). 
\end{equation}
where $pa(x_i)$  denotes the set of parent variables of  $x_i$. 
The value of each variable $x_i$ is computed as a linear combination of its parent variables $pa(x_i)$ and its corresponding noise variable $e_i$.
If coefficient $w_{ij}$ equals zero, this indicates that there is no direct causal effect from $x_j$ to $x_i (i, j=1,\cdots,u)$. 
Conversely, if the coefficient $w_{ij}$ is non-zero, it means that there is a direct causal effect from $x_j$ to $x_i$.
The noise terms $e_i (i = 1,\cdots,u)$ are independent and follow non-Gaussian continuous distributions~\cite{shimizu2006linear}.
Hence, the observed variables can be formulated
as $\boldsymbol{X} = \boldsymbol{W}\boldsymbol{X} + \boldsymbol{E}$,
where $\boldsymbol{E}$ is the noise matrix.
Therefore, $\boldsymbol{X}$ can be represented by:
\begin{equation}
\boldsymbol{X} = (\boldsymbol{I}-\boldsymbol{W})^{-1} \boldsymbol{E}.
\label{ica-problem}
\end{equation}
where $\boldsymbol{I}$ denotes the identity matrix.
Fortunately, equation~\eqref{ica-problem}
formulates an Independent Component Analysis (ICA) problem~\cite{hyvarinen2023nonlinear}, which is known to be identifiable~\cite{comon1994independent,shimizu2011directlingam}.
We use the ICA estimation method~\cite{shimizu2022statistical} to solve the ICA problem, yielding a resultant matrix denoted as $\boldsymbol{A}$.
The detailed solution process of the ICA problem can be found at~\cite{shimizu2006linear,shimizu2011directlingam}.
Finally, we can compute the causal strength matrix $\boldsymbol{W} = \boldsymbol{I}-\boldsymbol{A}^{-1}$.

\begin{figure*}[!t]
  \centering
\includegraphics[width=\linewidth]{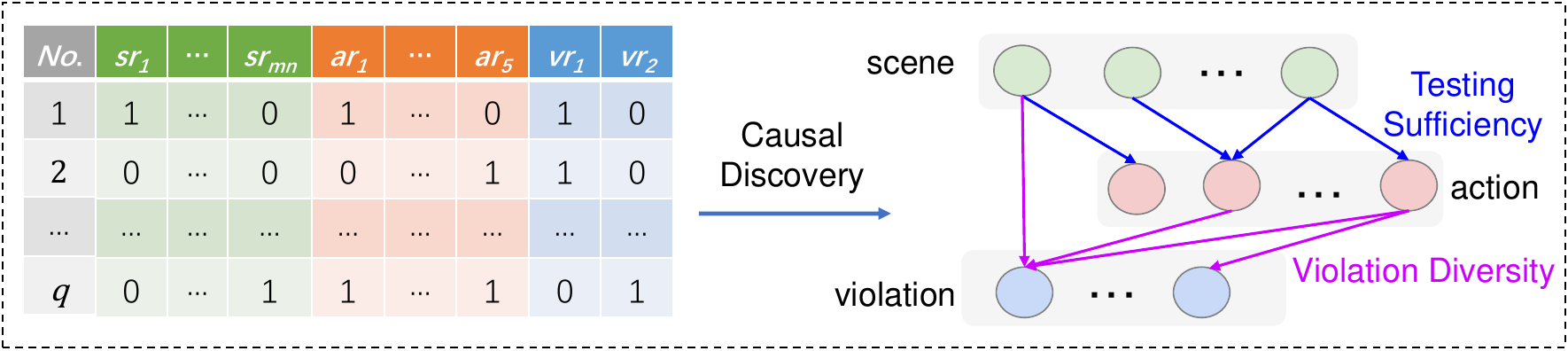}
  \caption{The process of discovering the scene-action-violation causal graph from a vectorized scenario representation.}
  \label{figure-causal-discovery}
\end{figure*}

Finally, we use  the constructed scenario representation dataset $\boldsymbol{X}_{sav} = [\boldsymbol{sr}_1, \cdots, \boldsymbol{sr}_{mn}; \boldsymbol{ar}_1, \cdots, \boldsymbol{ar}_{5};\boldsymbol{vr}_1,\boldsymbol{vr}_2]$ to extract the \textit{scene-action-violation} causal graph $\mathcal{G}_{sav}$.
The associated causal strength matrix is denoted as $\boldsymbol{W}_{sav}$.
The causal discovery process is shown in Fig.~\ref{figure-causal-discovery}.
Actually, the variables $scene$ and $action$  
correspond to the inputs and outputs of an ADS, where the scene influences the planning process that generates the appropriate actions for the vehicle~\cite{shao2024lmdrive,chen2024end}.
Hence, the edges among $scene \rightarrow action$ in $\mathcal{G}_{sav}$ represent the causal relationships activated during the planning process of the ADS within the current scenario.
If a testing process explores more \textit{scene-action} edges, it implies more comprehensive testing.
Consequently, the causal edges between $scene$ and $action$ can serve as a metric for measuring testing sufficiency.
For the edges among $(scene, action) \rightarrow violation $, the output represents the violation, while the inputs define the conditions (causes) under which the violation occurs.
Consequently, the edges between $(scene, action)$ and $violation$ can characterize the semantics of a violation.
Therefore, we can use these edges to comprehensively measure the diversity of violations.

\subsubsection{Feedback Calculation}
\label{feedback-computation}

To provide feedback to the fuzzer, we need to measure whether a new mutant $s^\prime$  can improve violation diversity or testing sufficiency in comparison to the existing test cases.
Given a causal graph $\mathcal{G}_{sav}$ and its corresponding causal strength matrix $\boldsymbol{W}_{sav}$,  an effective mutant  $s^\prime$  should introduce new edges or uncover novel combinations of existing edges within the causal graph.
Hence, we transform the weight matrix $\boldsymbol{W}_{sav}$ into a binary adjacency matrix  $\boldsymbol{B}_{sav}$.
In this binary matrix, if an element $w_{ij} = 1$, it indicates the presence of an edge from $x_j$ to $x_i$.
Given that the distance has been widely applied to measure diversity~\cite{cheng2023behavexplor,du2019deepstellar}, we calculate the minimum distance between binary matrices:
\begin{align}
    ts^{s^\prime} &= \min_{s \in \Delta} Distance (\boldsymbol{B}_{sa}^{s^\prime}, \boldsymbol{B}_{sa}^s), \\
    vd^{s^\prime} &= \min_{s \in \Delta} Distance ( \boldsymbol{B}_{sav}^{s^\prime}, \boldsymbol{B}_{sav}^s).
\end{align}
where $Distance (\cdot)$ computes the cosine similarity between two matrices.
The matrix $\boldsymbol{B}_{sa}^s$ is a submatrix of $\boldsymbol{B}_{sav}^s$, focusing exclusively on the causal edges from \textit{scene} to \textit{action}.
The values of $ts^{s^\prime}$ and $vd^{s^\prime}$ represent the testing sufficiency and violation diversity, respectively.
Therefore, a larger value of $ts^{s^\prime}$ indicates $s^\prime$ can improve the sufficiency of testing.
Similarly, a larger value of $vd^{s^\prime}$ signifies an increased diversity of violations within the scenario.
We configure two pre-defined thresholds $\theta_{ts}$ and $\theta_{cd}$, to evaluate whether there is a significant improvement in testing sufficiency and violation diversity.

In addition to providing diversity feedback, the primary target of ADS testing is to search for scenarios violating the given specifications. Therefore, we also need to assess the violation degree of a scenario, i.e., how far the scenario is from violating the safety specifications defined in \eqref{eq-safety-violation-sepecification}.
The violation degree of a seed $s$ is defined as follows:
\begin{equation}
    d^s = \sum_{v \in \{collision, destination\}} f_v (s).
\end{equation}
where $f_v(\cdot)$ is the violation degree with respect to the corresponding specification.
A lower value of $d^s$ indicates a better mutant.
As given in Section~\ref{preliminaries-ADS}, the criterion for a collision is defined as the minimum distance between the ego vehicle and NPC vehicles in the scenario being equal to 0.
Therefore, 
given the observation trace $T^s$ of a seed $s$,
we define the collision degree as:
\begin{equation}
    f_{collision} (s) = D_b(p^0, s).
\end{equation}
where $D_b(p^0, s)$ calculates 
 the minimum distance according to~\eqref{minimum-distance-calculation}.
In order to check whether the ego vehicle can finish the driving task, the violation degree of reaching the destination is defined by:
\begin{equation}
    f_{destination} (s) = \max 
    \left( 10-D_{p2p}(p_{t^s}^0, p_{dest}), 0
    \right).
\end{equation}
where $D_{p2p}(p_{t^s}^0, p_{dest})$ calculates the distance between the last
position of the ego vehicle (i.e., $p_{t^s}^0$) and the destination $p_{dest}$.

\subsection{Causality-driven Mutation}

\label{causal-mutation}

\begin{figure}[!t]
  \centering
\includegraphics[width=\linewidth]{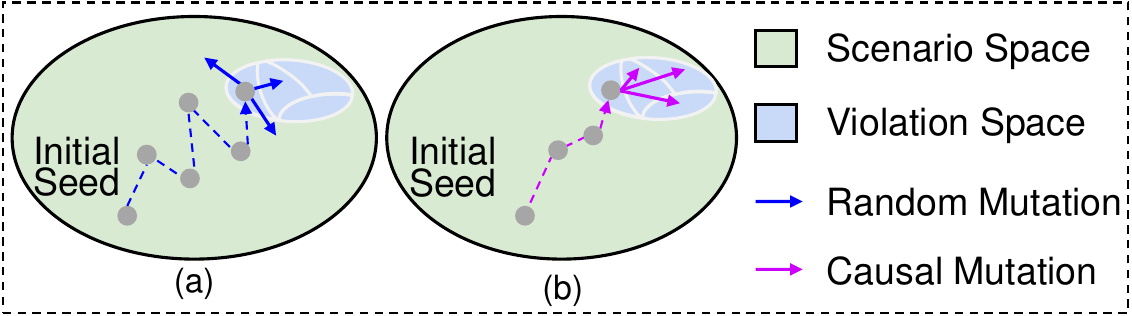}
  \caption{Illustration of (a) existing random mutation and (b) our causal-driven adaptive mutation.}
  \label{figure-causal-mutation-comparision}
\end{figure}

Although existing methods typically generate new test scenarios through random mutations, the inherent scarcity of violation-inducing scenarios in the scenario space renders this random mutation strategy inefficient for multi-objective search~\cite{fu2024icsfuzz}.
Intuitively, randomness can thoroughly explore the ADS violation space from different directions originating from the safety space. 
However,
the violation space only occupies a small part of the entire scenario space~\cite{zhong2022neural}.  
Especially, as shown in Fig.~\ref{figure-causal-mutation-comparision}, once a violation scenario is identified, we should mutate 
along specified directions
instead of randomly exploring, as random mutation may cause the search process to jump out of the violation space (Fig.~\ref{figure-causal-mutation-comparision}(a)).
However, as shown in Fig.~\ref{figure-causal-mutation-comparision}(b), a well-designed mutation strategy, coupled with fine-grained optimization, can effectively guide the fuzzing process to explore diverse scenarios within the violation space.
To mitigate this challenge, we propose a causality-driven mutation strategy, which employs an interpretable method to identify the key objects for mutation.
Specifically, it first qualitatively estimates the causal effect (i.e., contribution) of each NPC vehicle on changes in ego actions and the occurrence of violation.
Then, based on the estimated causal effects, we design an adaptive mutation strategy that encourages the mutation process to prioritize objects with a greater causal effect.

\subsubsection{Causal Effect Estimation}
The edges in a causal graph indicate the existence of causal relationships between variables, whereas causal effects provide a quantitative measure of the strength of these relationships.
Hence, in this section, given a scenario $s$ with an ego vehicle and $ n^{NPC}$ NPC vehicles, we aim to estimate the causal effect of each NPC vehicle on the ego vehicle, i.e., $ACE(npc_k \rightarrow ego), k = 1, \cdots,  n^{NPC}$.
Formally, the causal effect can be described as:
\begin{equation}
    ACE(npc_k \rightarrow ego) = \frac{1}{t^s}\sum_{t=1}^{t^s} [\boldsymbol{sr}_t(k)] \times [w_{1}, \cdots, w_{mn}]^\mathrm{T}.
    \label{eq-causal-effect-npc-ego}
\end{equation}
where vector $\boldsymbol{sr}_t(k)$ is the scene representation with only NPC vehicle $k$.
Moreover, the weight $w_i = \sum_{j=1}^{p} w_{ij}, i = 1, \cdots, p$, denotes the sum of causal strengths of all edges that  going out from the $i$-th variable, where 
$w_{ij}$ is the weight in 
the causal strength matrix (i.e., $\boldsymbol{W}_{sav}$) of $\mathcal{G}_{sav}$, which was discovered in Section~\ref{causal-graph-discovery}.

\begin{figure*}[!t]
  \centering
\includegraphics[width=1\linewidth]{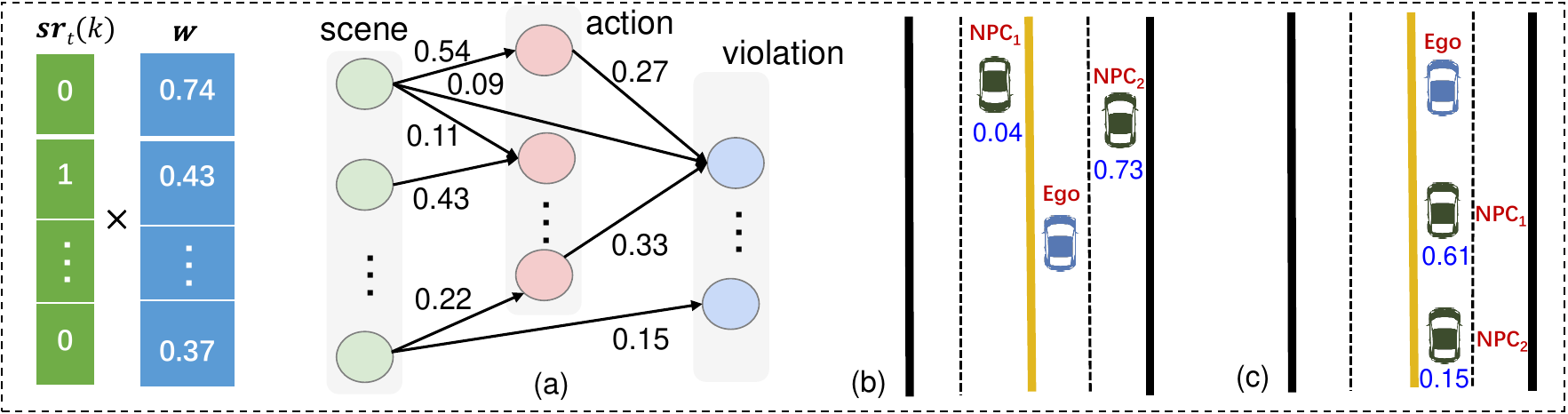}
  \caption{(a) The calculation process of the causal effect; (b) A scenario with two NPC vehicles on lanes moving in opposite directions; (c) A scenario with two NPC vehicles on lanes moving in the same direction.
 The causal effect of each NPC vehicle is marked in blue.
  }
  \label{figure-npc-causal-effect}
\end{figure*}

To illustrate the main idea of causal effect estimation for each NPC, we provide an example, as shown in Fig.~\ref{figure-npc-causal-effect}(a), assuming $\boldsymbol{sr}_t(k) = [0,1,\cdots,0]$.
Based on the causal strength matrix $\boldsymbol{W}_{sav}$, we can obtain the causal strength of each node, i.e., $[0.74, 0.43, \cdots, 0.37]$.
For example, for the first \textit{scene} node, its causal effect can be calculated as $0.54 + 0.09 + 0.11 = 0.74$.
Hence, the causal effect of the NPC vehicle on the ego vehicle at the current time instant can be calculated as $(0\times0.74) + 
(1\times0.43)+
\cdots + (0\times0.37)$.
Finally, after obtaining the causal effect for each time instant, we can calculate the average causal effect of each NPC vehicle, according to~\eqref{eq-causal-effect-npc-ego}.

\subsubsection{Adaptive Mutation}
Once the causal effect of each NPC vehicle is estimated, this information can be leveraged to guide the mutation process.
Specifically, the causal information helps us distinguish causal (risk) NPC vehicles, which have a significant influence on the ego vehicle, from non-causal NPC vehicles, which have minor or no influence.
Consequently, a mutation strategy can focus more on causal instances to efficiently generate potential critical scenarios.

As shown in Fig.~\ref{figure-npc-causal-effect}(b), the scenario contains an ego vehicle and two NPC vehicles, i.e., $NPC_1$ and $NPC_2$. 
According to the results of the causal effect estimation, their causal effects are $0.04$ and $0.73$, respectively.
This indicates that $NPC_2$ significantly influences the ego vehicle, while $NPC_1$ has a minimal effect, as it is driving in the opposite lane and does not affect the ego vehicle's behavior. In contrast, $NPC_2$ is in the adjacent lane moving in the same direction, thus impacting the ego vehicle more directly.
Fig.~\ref{figure-npc-causal-effect}(c) shows another scenario, where two NPCs are driving behind the ego vehicle, with causal effects of $0.61$ and $0.15$, respectively.
 The higher causal effect of $NPC_1$ is attributed to its closer proximity to the ego vehicle, allowing it to influence the ego's behavior directly. 
However, $NPC_2$ has a diminished impact because it is separated by $NPC_1$.

Algorithm~\ref{alg2} shows the adaptive mutation process. It employs an $\epsilon$-greedy strategy~\cite{ding2023incremental} to balance exploration and exploitation.
When a randomly sampled value is less than the exploration probability  $\epsilon$, a mutant $s^\prime$ is generated through random mutation (Lines~\ref{alg-random-mutation-start}--\ref{alg-random-mutation-end}). Otherwise, in Lines~\ref{alg-causal-mutation-start}--\ref{alg-causal-mutation-end}, 
based on the estimated causal effects of each NPC vehicle, different weights are assigned, thereby increasing the probability of mutating key causal instances.
For each NPC vehicle, we perform waypoint mutation~\cite{cheng2023behavexplor}.
For each waypoint, we modify its velocity and position, as described in Section~\ref{preliminaries-ADS}.
Finally, a mutant $s^\prime$ is generated (Line~\ref{causal-mutant-generated}).

\begin{algorithm}[!t]
\caption{Causal-driven Adaptive Mutation Algorithm.}
\label{alg2}
\begin{algorithmic}[1]
    \REQUIRE Selected scenario seed $s$, the exploration
probability $\epsilon$, the estimated causal effect of each NPC vehicle $ACE_k$.
    \ENSURE Mutated scenario seed $s^\prime$.
   \IF{$\texttt{Random()} < \epsilon$} \label{alg-random-mutation-start}
   \STATE $s^\prime \leftarrow \texttt{RandomMutation}(s)$; \\ \label{alg-random-mutation-end}
   \ELSE
   \FOR{$k \in \{1, \cdots, n^{NPC}\}$}
   \label{alg-causal-mutation-start}
   \STATE Mutate NPC vehicle $k$ with an adaptive probability of $ACE_k/\sum_{i=1}^{n^{NPC}}ACE_i$;\\
    \label{alg-causal-mutation-end}
   \ENDFOR
   \STATE $s^\prime \leftarrow \texttt{UpdateScenario}(s)$; \\
   \label{causal-mutant-generated}
   \ENDIF
\RETURN  $s^\prime$.
\end{algorithmic}
\end{algorithm}

\section{Evaluation}
\label{experiment-sec}
In this section, we evaluate the effectiveness and efficiency
 of \method~in generating critical scenarios. We aim to answer the following research questions:
\begin{itemize}[leftmargin=*]
\item 
RQ1: Can \method~correctly learn the interrelationships among scenes, actions, and
violations in a scenario?
\item 
RQ2: Can \method~ effectively discover diverse violations and ensure high testing sufficiency in comparison to the selected baselines?
\item
RQ3: How efficient is \method~in terms of computation time and the number of scenarios explored to discover the first critical scenario?
\item
RQ4: How useful are the causality-based feedback and the causality-driven mutation in improving testing performance?
\end{itemize}

\subsection{Experiment Setup}
\begin{figure*}[!t]
  \centering
\includegraphics[width=\linewidth]{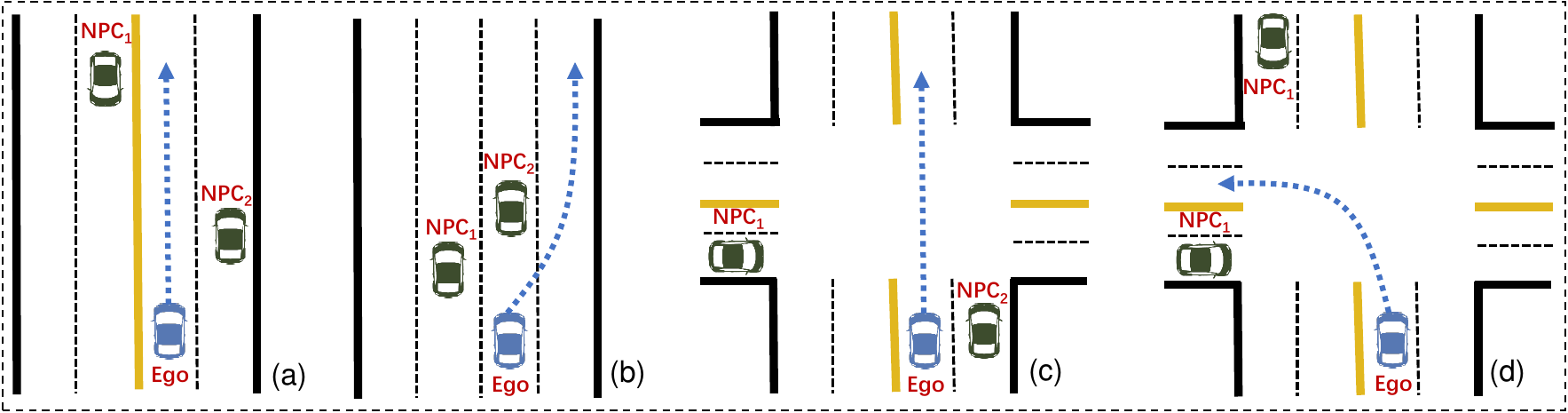}
  \caption{Four functional scenarios for testing.
  }
  \label{figure-four-scenarios}
\end{figure*}

To answer the four research questions, we chose Baidu Apollo as the target ADS because of its status as one of the most prominent open-source ADS platforms, with extensive commercialization and recognition in numerous ADS-related studies~\cite{cheng2023behavexplor,li2024viohawk,huai2023doppelganger,sun2024redriver}. Specifically, we use  Apollo 7.0 for the following experiments.
We use the LGSVL simulator~\cite{rong2020lgsvl} to simulate real-world
environment and vehicle dynamics.
In particular, we use SORA-SVL~\cite{Sorasvl2023} to establish the connection between the simulator and Apollo.
We run \method~on the following four functional scenarios related to straight roads and intersections as they represent common real-world driving tasks and have been extensively tested in the literature~\cite{cheng2023behavexplor,huai2023doppelganger,li2024viohawk}.
$Scenario1$: As shown in Fig.~\ref{figure-four-scenarios}(a), the ego vehicle follows a lane at a straight road with two-way four lanes.
$Scenario2$: As shown in Fig.~\ref{figure-four-scenarios}(b), the ego vehicle changes lanes on a straight road with one-way four lanes.
$Scenario3$: As shown in Fig.~\ref{figure-four-scenarios}(c), the ego vehicle goes straight through a non-signalized intersection.
$Scenario4$: As shown in Fig.~\ref{figure-four-scenarios}(d), the ego vehicle turns left at a non-signalized intersection.
All scenarios are generated using the San Francisco map.

We compare \method~with three representative techniques:
random testing, AV-Fuzzer~\cite{li2020av}, and DoppelTest~\cite{huai2023doppelganger}.
Specifically, random testing operates without feedback, selecting and mutating test cases randomly from the seed corpus. AV-Fuzzer is a two-phase approach that first applies a genetic algorithm to explore scenarios with a high likelihood of violating safety requirements, followed by a local fuzzer to search for safety violations within the generated high-risk scenarios. DoppelTest employs a genetic algorithm to uncover bug-revealing violations by orchestrating multiple instances of the same ADS.
For fair comparisons, we only consider vehicle-vehicle collisions in DoppelTest and do not add pedestrians to the traffic flow.
In the experiments, we set $m = 8$ and $n = 4$, resulting in $8 \times 4 = 32$ areas in the scene partition.
Similar to~\cite{cheng2023behavexplor,li2020av}, the threshold $\theta_p$ in \eqref{eq-safety-violation-sepecification} is set as 1.
The parameters $\theta_{ts}$, $\theta_{vd}$, $\theta_a^+$, $\theta_a^-$, $\theta_h$ are empirically set to $0.3$, $0$, $0.1$, $-0.1$, $0.1$, respectively.
The exploration rate of $\epsilon$-greedy strategy is 0.5.
The perception range $L$ is set to 50.

\subsection{RQ1: Correctness of the Discovered Causal Graphs}
In this section, we evaluate the correctness of the discovered causal graph
to determine whether it effectively captures the essential causal relationships within a given scenario.
We conduct experiments using 200 scenarios, generating a corresponding causal graph for each scenario, which results in a total of 200 causal graphs.
We evaluate these causal graphs using DoWhy\footnote{\url{https://www.microsoft.com/en-us/research/project/dowhy/}}, a toolbox developed by Microsoft.

Table~\ref{tab1-causal-graph} shows the evaluation results of the causal structure, causal mechanisms, and causal effects.
For the causal graph structure, we conduct the permutation-based test~\cite{eulig2023toward} and the conditional independence test~\cite{zanga2022survey} to evaluate both global and local Markov equivalence between the causal graph and the observed scenario data.
As shown in Table~\ref{tab1-causal-graph}, based on the mean and variance of the significance level (i.e., $p$-value), we observe that the $p$-value is consistently less than 0.05, indicating that the discovered causal graphs can accurately represent the causal structure present in the observational data.
For the causal mechanism, we calculate the Kullback–Leibler (KL) divergence between the generated and observed distributions, as well as the Mean Squared Error (MSE) between the observational data and the causal graph predictions. These two metrics evaluate whether the discovered causal graphs effectively capture the underlying data-generation mechanism, with KL divergence applied to root nodes and MSE to non-root nodes.
As shown in Table~\ref{tab1-causal-graph}, the results indicate that the discovered causal graphs accurately capture the causal mechanisms embedded in the observed scenario data.
For causal effect, we evaluate its robustness using a series of refutation tests~\cite{sharma2020dowhy}. Given that causality should remain invariant, changes in the data should not alter the estimation of causal effects. Thus, if a causal effect estimation fails the refutation test (i.e., $p$-value $<$ 0.05), it indicates a potential issue with the estimation. We perform the refutation test using two methods: adding a random common cause variable and removing a random subset of the data. 
From the results of Table~\ref{tab1-causal-graph},
we can find that the causal effect estimation exhibits reasonable robustness to refutation.

\begin{table}
  \centering
  \tabcolsep 1pt
  \caption{Evaluation Results of the Discovered Causal Graphs}
    \begin{tabular}{c|cc|cc|cc}
   \hline
    \multicolumn{1}{c|}{\multirow{2}[0]{*}{Metric}} & \multicolumn{2}{c|}{Causal Graph Structure} & \multicolumn{2}{c|}{Causal Mechanism} & \multicolumn{2}{c}{Causal Effect} \\
    \cline{2-7}
          & \multicolumn{1}{p{5em}}
          {\centering Permutation-Based Test} & \multicolumn{1}{p{6em}|}{\centering Independence Test} & 
          \multicolumn{1}{p{4em}}{\centering KL  Divergence} & \multicolumn{1}{p{2em}|}{\centering MSE} & \multicolumn{1}{p{3.5em}}{\centering Common Cause} & \multicolumn{1}{p{3.5em}}{\centering Random Subset} \\
    \hline
    Mean  & 0.001  & 0.001  & 1.423  & 0.112  & 0.918  & 0.869  \\
    Variance & 0.001  & 0.001  & 0.798  & 0.001  & 0.004  & 0.012  \\
    \hline
    \end{tabular}%
  \label{tab1-causal-graph}%
\end{table}



\begin{figure}[!t]
  \centering
  \begin{minipage}[t]{0.49\linewidth} 
    \centering
    \includegraphics[width=\linewidth]{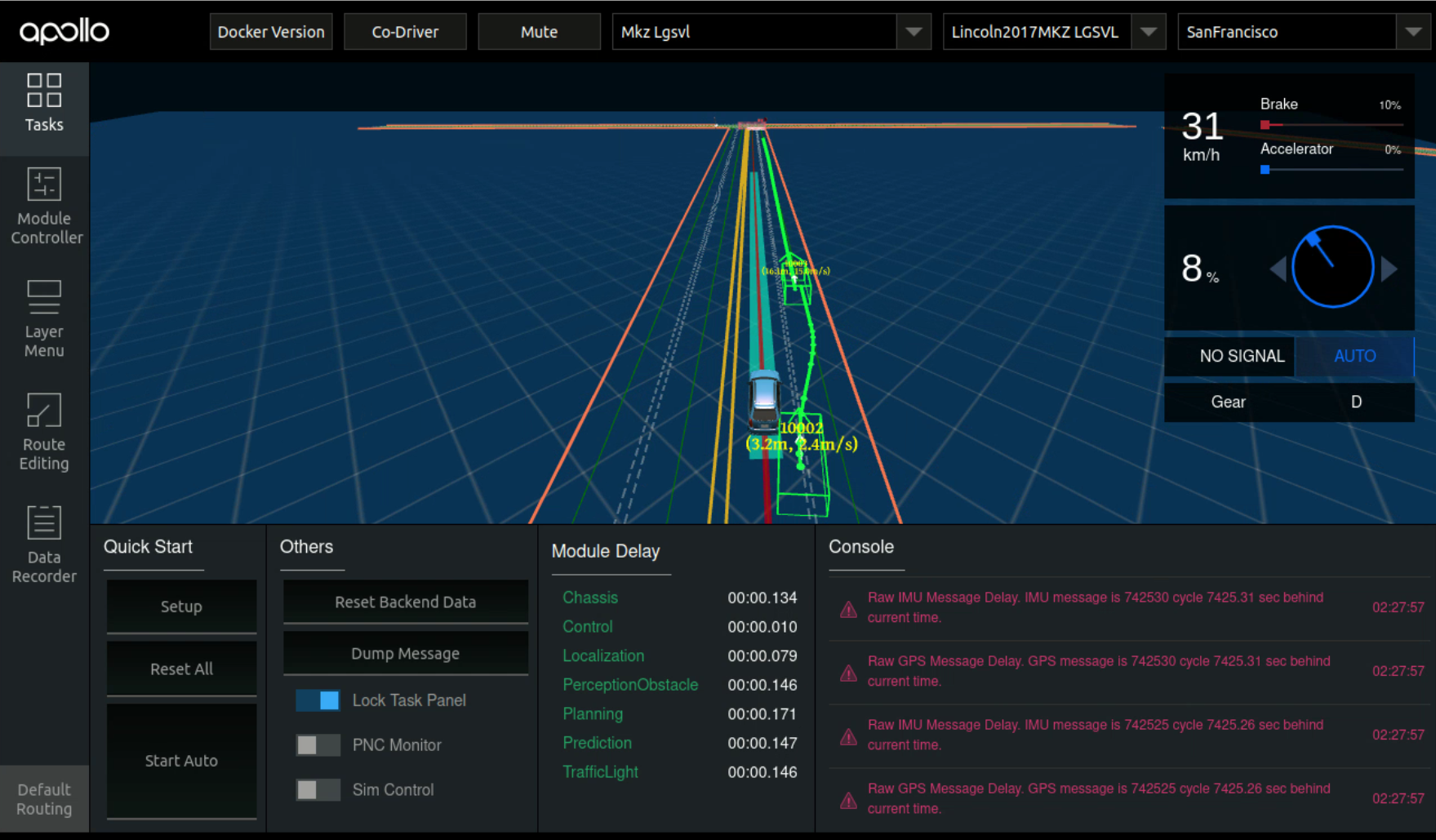}
    \vspace{0.5em}
    \centering (a) Scenario
  \end{minipage}
  \hfill 
  \begin{minipage}[t]{0.49\linewidth}
    \centering
    \includegraphics[width=\linewidth]{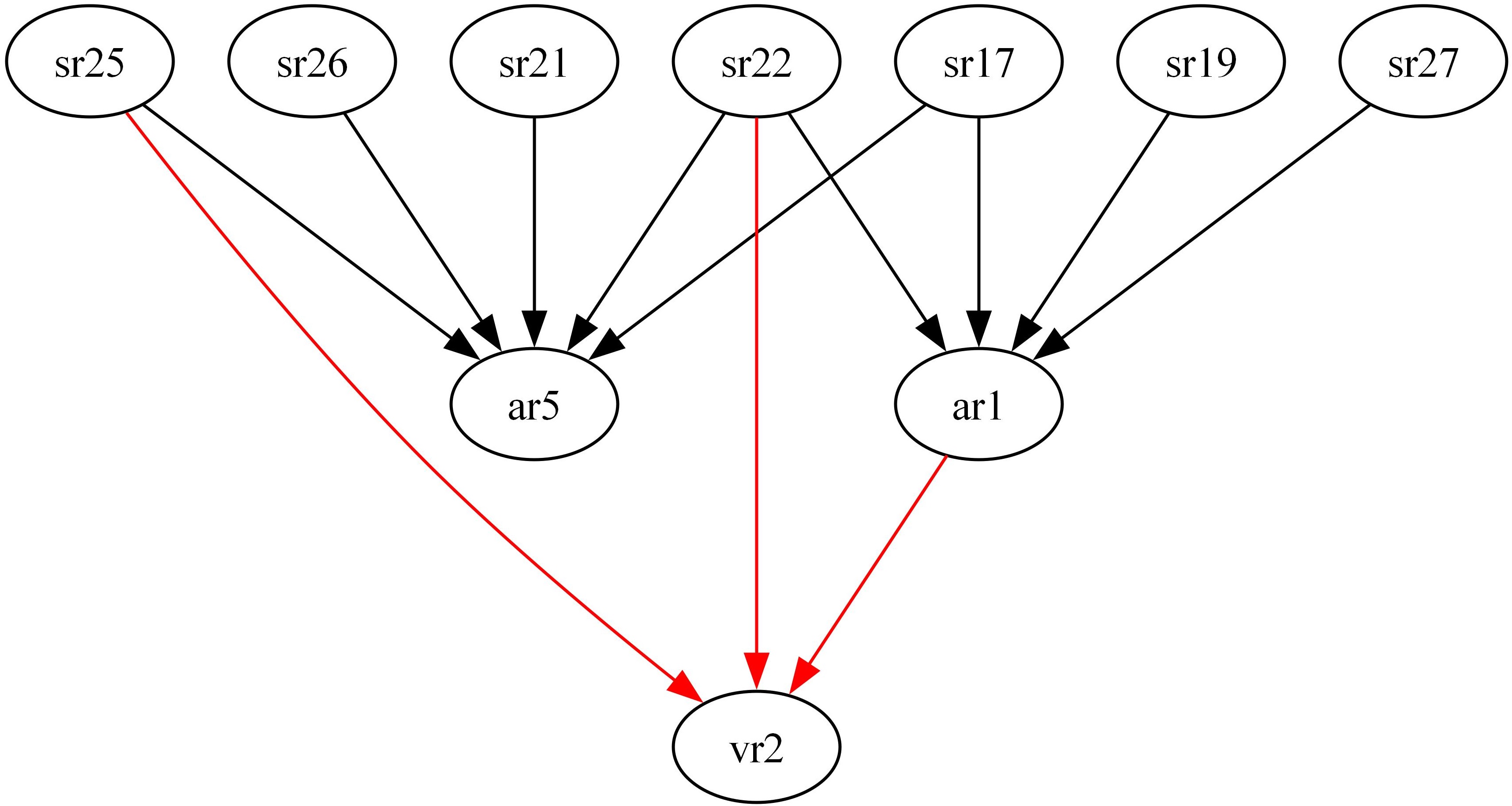}
    \vspace{0.5em}
    \centering (b) Causal graph
  \end{minipage}
  \caption{A collided scenario where the collision is caused by an NPC vehicle and its causal graph.}
  \label{case-study-scenario-graph}
\end{figure}

Figure~\ref{case-study-scenario-graph}(a) shows a snapshot of Apollo Dreamview, where a scenario with one ego vehicle and two NPC vehicles is executed. We can observe that at the current moment, a collision occurs between the ego and $NPC_1$. Based on the collected observation trace, we construct a causal graph, as shown in Fig.~\ref{case-study-scenario-graph}(b).
In the graph, we observe a $(scene, action) \rightarrow violation$ causal edge combination, i.e., $(sr_{25}, sr_{22}; ar_1) \rightarrow vr_2$, which indicates that a \textit{violation} ($vr_2$, an NPC-induced collision) 
has occurred in a \textit{scene} where two NPCs have been positioned at the right front ($sr_{25}$) and right rear ($sr_{22}$), respectively, 
while the ego vehicle has executed an accelerated \textit{action} ($ar_1$).
This causal graph represents a category of collision scenarios in which the ego vehicle is positioned between two NPC vehicles, one at the right front and the other at the right rear. In this scenario, the ego vehicle attempts to accelerate for safety, but the trailing NPC takes an inappropriate action, leading to a collision.
Thus, the causal graph semantically corresponds to the scenario observation in Fig.~\ref{case-study-scenario-graph}(a). 

\subsection{RQ2: Effectiveness of \method}
In this section, we evaluate the capability of \method~to discover diverse violations. We compare \method~with three baseline methods under four functional scenarios, running each method twice per scenario type for a duration of 10 hours each time.
Table~\ref{tab2-violation-discovery} shows the comparison results,  with each value representing the average result of two runs, totaling 20 hours.
From the results, we can see that within the same amount of time, \method~can discover more violations across all four functional scenarios.
For Scenarios 1–4, the numbers of discovered 
 violations are $110$, $67$, $148.5$, and $68$, respectively.
This means that \method~can effectively find more violations than the existing baseline methods.
Furthermore, for the testing sufficiency metric, \method~successfully explored the highest number of SAC in three out of the four scenarios (i.e., scenarios 2, 3, and 4).
For scenario 1, although random testing explored more SAC, the number of violations is significantly lower than those discovered by \method~(7 vs. 110).
The results demonstrate that \method~achieves a better trade-off between violation detection and testing sufficiency.
In addition, we can also observe that \method~exhibits better performance in terms of SAVC across all scenarios compared to the three baselines.
The reason is that the incorporation of causal feedback and violation degree can effectively guide the search process of critical scenarios.
As shown in the last three rows of Table~\ref{tab2-violation-discovery}, \method~significantly outperforms the three baselines across all metrics in the average results, discovering 56 extra violations, enhancing testing sufficiency by 22.06\%, and increasing violation diversity by 134.09\% compared to the best-performing baseline.

\begin{table}
  \centering
  \tabcolsep 4pt
  \caption{Effectiveness Comparison Results with Three Baselines. SAC: The Number of Unique Combinations of \textit{scene-action} Causal Edges, SAVC: The Number of Unique Combinations of \textit{scene-action-violation} Causal Edges}
    \begin{tabular}{c|c|cccc}
    \hline
   \multicolumn{1}{c|}{\multirow{2}{*}{Scenario}} & \multicolumn{1}{c|}{\multirow{2}{*}{Metric}} & \multicolumn{4}{c}{Method}                      \\\cline{3-6}
\multicolumn{1}{c|}{}                          & \multicolumn{1}{c|}{}                         & Random & AV-Fuzzer & DoppelTest & Causal-Fuzzer \\
    \hline
    \multirow{3}[0]{*}{1} & 
    Violation & 7.0     & 39.5  & 77.0    & \textbf{110.0} \\
          & SAC   & \textbf{181.0}   & 143.5 & 150.5 & 155.5 \\
          & SAVC  & 7.0     & 7.0     & 7.0    & \textbf{11.0} \\
          \hline
    \multirow{3}[0]{*}{2} & Violation & 3.0     & 34.5  & 55.0    & \textbf{67.0} \\
          & SAC   & 93.0    & 110.0   & 78.0    &  \textbf{150.0} \\
          & SAVC  & 3.0     & 4.5   & 3.0     & \textbf{13.0} \\
         \hline
    \multirow{3}[0]{*}{3} & Violation & 16.5  & 53.0    & 10.0    &  \textbf{148.5} \\
          & SAC   & 114.0   & 155.5 & 155.0   & \textbf{167.0} \\
          & SAVC  & 9.0     & 12.0    & 8.0     & \textbf{17.5} \\
          \hline
    \multirow{3}[0]{*}{4} & Violation & 3.5   & 18.5  & 27.5  & \textbf{68.0} \\
          & SAC   & 56.5  & 121.5 & 217   & \textbf{260.5} \\
          & SAVC  & 3.5   & 3.5   & 4.0     & \textbf{10.0} \\
           \hline
    \multirow{3}[0]{*}{Average} & Violation & 7.5   & 36.4  & 42.4  & \textbf{98.4}  \\
          & SAC   & 111.1  & 132.6  & 150.1  & \textbf{183.3}  \\
          & SAVC  & 5.6   & 6.8   & 5.5   & \textbf{12.9}  \\
          \hline
    \end{tabular}%
  \label{tab2-violation-discovery}%
\end{table}%


\subsection{RQ3: Efficiency
 of \method} 
In this section, we evaluate the efficiency of \method~from
different perspectives.
We first calculate the mutation time and feedback time to quantify the time required for generating a mutate and computing a feedback. To evaluate the efficiency of discovering critical scenarios, we also record the number of scenarios explored to
discover the first critical scenario.
The results are given in Table~\ref{tab3-efficiency}.
From the results, we can observe that there is no significant difference in mutation time among the four methods.
This means that \method~does not cause an increase in time during the mutation phase.
During the feedback phase, since \method~needs to construct a causal graph (i.e., a step involving graph searching), it requires more time.
Hence, as shown in Table~\ref{tab3-efficiency}, the average feedback time for \method~is 4.313 seconds, while the feedback times of the three baseline methods are 0.040, 0.060, and 4.928 seconds, respectively.
This means that compared to random testing and AV-Fuzzer, \method~experiences an increase in time consumption during the mutation phase. However, considering that \method~can discover more violations within the same amount of time, this indicates that the increase in feedback time does not negatively impact its violation detection capability.
For DoppelTest, more oracles, such as module response failure, lead to higher feedback time, with an average of 4.928 seconds.

Furthermore, by comparing the number of scenarios explored to find the first critical scenario, we can observe that, on average, \method~triggers the first critical scenario after exploring 32.125 scenarios.
For Scenario 2, we observe that compared with AV-Fuzzer, more scenarios are required to detect the first critical scenario.
The reason is that the presence of multiple one-way lanes in Scenario 2, where all NPC vehicles influence the behavior of the ego vehicle, leads to minimal differentiation in their causal effects on the ego during the initial phase.
However, as time progresses and additional causal edges are incorporated, \method~is able to discover the most violations.
Thus, we can conclude that by integrating fine-grained causal effect estimation, \method~enhances mutation efficiency and accelerates the search process for critical scenarios.

\begin{table}
  \centering
    \tabcolsep 1.6pt
  \caption{Efficiency Comparison Results of \method~and Three Baselines}
    \begin{tabular}{c|c|cccc}
    \hline
     \multicolumn{1}{c|}{\multirow{2}{*}{Scenario}} & \multicolumn{1}{c|}{\multirow{2}{*}{Metric}} & \multicolumn{4}{c}{Method}                      \\\cline{3-6}
\multicolumn{1}{c|}{}                          & \multicolumn{1}{c|}{}                         & Random & AV-Fuzzer & DoppelTest & Causal-Fuzzer \\
\hline
    \multirow{3}[0]{*}{1} & Mutation Time & \multicolumn{1}{c}{4.259E-06} & \multicolumn{1}{c}{2.421E-03} & \multicolumn{1}{c}{3.246E-03} & \multicolumn{1}{c}{1.881E-03} \\
          & Feedback Time & \multicolumn{1}{c}{0.073 } & \multicolumn{1}{c}{0.073 } & \multicolumn{1}{c}{3.845 } & \multicolumn{1}{c}{3.896 } \\
          & First Failure & \multicolumn{1}{c}{164.500 } & \multicolumn{1}{c}{156.500 } & \multicolumn{1}{c}{147.000 } &  \multicolumn{1}{c}{ \textbf{7.500} } \\
          \hline
    \multirow{3}[0]{*}{2} & Mutation Time & \multicolumn{1}{c}{6.997E-04} & \multicolumn{1}{c}{1.370E-03} & \multicolumn{1}{c}{5.206E-04} & \multicolumn{1}{c}{2.040E-03} \\
          & Feedback Time & \multicolumn{1}{c}{0.034 } & \multicolumn{1}{c}{0.049 } & \multicolumn{1}{c}{5.985 } & \multicolumn{1}{c}{4.545 } \\
          & First Failure & \multicolumn{1}{c}{31.000 } & \multicolumn{1}{c}{\textbf{27.000} } & \multicolumn{1}{c}{73.000 } & \multicolumn{1}{c}{69.000 } \\
         \hline
    \multirow{3}[0]{*}{3} & Mutation Time & \multicolumn{1}{c}{1.517E-06} & \multicolumn{1}{c}{1.815E-03} & \multicolumn{1}{c}{3.012E-03} & \multicolumn{1}{c}{2.497E-03} \\
          & Feedback Time & \multicolumn{1}{c}{0.033 } & \multicolumn{1}{c}{0.059 } & \multicolumn{1}{c}{4.829 } & \multicolumn{1}{c}{4.736 } \\
          & First Failure & \multicolumn{1}{c}{37.500 } & \multicolumn{1}{c}{25.000 } & \multicolumn{1}{c}{9.000 } & \multicolumn{1}{c}{\textbf{5.500} } \\
          \hline
    \multirow{3}[0]{*}{4} & Mutation Time & \multicolumn{1}{c}{3.829E-06} & \multicolumn{1}{c}{1.803E-03} & \multicolumn{1}{c}{4.791E-04} & \multicolumn{1}{c}{1.982E-03} \\
          & Feedback Time & \multicolumn{1}{c}{0.019 } & \multicolumn{1}{c}{0.060 } & \multicolumn{1}{c}{5.052 } & \multicolumn{1}{c}{4.077 } \\
          & First Failure & \multicolumn{1}{c}{82.000 } & \multicolumn{1}{c}{165.000 } & \multicolumn{1}{c}{57.000 } & \multicolumn{1}{c}{ \textbf{46.500} } \\
    \hline
    \multirow{3}[0]{*}{Average} & Mutation Time & 1.773E-04 & 1.852E-03 & 1.814E-03 & 2.10E-03\\
          & Feedback Time & 0.040  & 0.060  & 4.928  & 4.313  \\
          & First Failure & 78.750  & 93.375  & 71.500  &  \textbf{32.125}  \\
          \hline
    \end{tabular}%
  \label{tab3-efficiency}%
\end{table}%

\subsection{RQ4: Ablation Study}

Finally, we conduct an ablation study to justify the contribution of each key component in \method. By isolating each component, we analyze its specific impact on overall performance.
Using \method~as the baseline, we design two variants: Variant 1, which excludes the causality-based feedback module, and Variant 2, which excludes the causality-driven mutation module.
Specifically, in Variant 1, we remove the causality-based feedback, retaining only the violation degree as feedback. In Variant 2, we replace the causality-driven mutation with a random mutation.
The results are shown in Table~\ref{tab4-ablation-study}.

\textbf{Usefulness of Causality-driven Mutation.} 
Considering the third row in Table~\ref{tab4-ablation-study}, we observe a decline in model performance when the causal feedback module is removed. However, with the aid of causality-driven mutation, Variant 1 still discovers a relatively high number of SAVC, even though the test sufficiency (i.e., the number of SAC) significantly decreases. This result indicates that integrating causality into the mutation process enhances the efficiency of diverse violation discovery.

\textbf{Usefulness of Causality-based Feedback.}
Considering the last row in Table 4, once causal-driven mutation is removed, the number of discovered SAVC drops significantly, from 12.875 to 7.875, while the reduction in SAV is smaller, decreasing from 183.250 to 178.500. 
This means that removing the causal-driven mutation module will significantly
reduce the diversity of violations.
The results of SAV indicate that incorporating causality-based feedback can significantly enhance testing sufficiency, thereby improving the comprehensiveness of the testing process.
Furthermore, by comparing the results of \method~and its variants, we can find that integrating causality-driven mutation and causality-based feedback within a unified framework enables \method~to achieve superior performance.
These findings from the ablation study validate the design of \method.

\begin{table}
 \tabcolsep 15pt
  \caption{Results of Ablation Study}
  \label{tab4-ablation-study}
  \begin{tabular}{c|ccc}
     \hline
    Method & Violation & SAC  & SAVC  \\
     \hline
    Causal-Fuzzer & \textbf{98.375} & \textbf{183.250} &\textbf{12.875}  \\
     Variant1 & 71.750 & 123.750 & 11.000     \\
     Variant2 & 52.625 & 178.500 & 7.875 \\
      \hline
\end{tabular}
\end{table}


\section{Related Work}
\label{related-work-sec}

\subsection{Scenario-based ADS testing}
Scenario-based testing is widely recognized for evaluating ADS reliability, as it enables the efficient generation of configurable, critical, and representative scenarios.
Several techniques have been proposed by researchers. These works can be categorized into data-driven methods~\cite{tang2023survey,zhang2023building,guo2024sovar,deng2023target} and search-based methods~\cite{li2024viohawk,kim2022drivefuzz,cheng2023behavexplor,he2024curiosity,biagiola2024boundary,tian2022mosat,wang2025moditector,cheng2024decictor,tang2025moral}.
Data-driven methods aim to generate critical scenarios from
existing scenario description resources, such as accident reports~\cite{guo2024sovar,song2025synthetic} and traffic rules~\cite{deng2023target}.
Search-based methods aim to search for critical scenarios using different technologies, such as evolutionary algorithms~\cite{zhong2022neural,li2020av,huai2023sceno} and reinforcement learning~\cite{lu2022learning,haq2023many}.
These existing methods primarily focus on generating violation scenarios to expose potential vulnerabilities of ADS.
Ensuring scenario diversity and testing sufficiency remains a challenge for search-based methods, prompting some recent research efforts to mitigate this issue.
For example, Cheng et al.~\cite{cheng2023behavexplor} propose behavior guidance to enhance the behavior diversity of the ego vehicle.
In~\cite{hildebrandt2023physcov}, testing sufficiency is evaluated based on the physical spatial areas covered during the testing process.
However, a fundamental limitation of existing methods is their focus on isolated diversity metrics rather than a comprehensive diversity that accounts for interrelationships among different factors. Furthermore, they do not effectively ensure both violation diversity and testing sufficiency simultaneously.

\subsection{Causality for Testing}
Causality is the study of how variables influence one another and how causes lead to effects~\cite{pearl2009causality}.
Formalizing, identifying, and quantifying causality within a system is essential for understanding and reasoning about its complex mechanisms.
Recently, causality has become a widely adopted methodology for analyzing complex systems~\cite{sun2024neural,yan2024causality,li2023droid}.
For example, causality has been applied to identify risk factors, enhancing the efficiency of test case generation~\cite{jiang2024generation}.
In~\cite{causalaf2022}, the causal graph obtained from human prior knowledge is incorporated into the scenario generation model.
In~\cite{giamattei2024causality}, the authors extracted causal relationships as a surrogate model, which helps to 
filter test cases and 
reduce the exploration required across the search space.
In addition, causality has been applied as a tool to analyze the causal events of an accident~\cite{sun2024acav}.
Different from these works, we utilize causality to measure violation diversity and guide the mutation process.

\section{Conclusion}
\label{conclusion-sec}
In this paper, we propose \method, a novel testing method designed to generate critical scenarios while simultaneously ensuring both violation diversity and testing sufficiency.
To measure the violation diversity and testing sufficiency, we extract a \textit{scene-action-violation} causal graph to formalize the interrelationships between input scenarios,
ADS actions, and violations.
Based on the edges in the causal graph, we propose a causality-based feedback method to quantify the violation diversity and testing sufficiency metrics, incorporating these two metrics into the testing framework.
Based on the strengths in the causal graph, we propose a causality-driven adaptive mutation strategy to facilitate the search process for critical scenarios.
Evaluation results on Apollo
demonstrated the effectiveness and efficiency of our method.
In future work, we aim to integrate additional semantic information into causal graphs to enhance diversity evaluation. Additionally, we plan to explore causality-based testing for other autonomous intelligent systems.

\bibliographystyle{IEEEtran}
\bibliography{ref}

\end{document}